# Short channel effects in graphene-based field effect transistors targeting radio-frequency applications

Pedro C Feijoo, David Jiménez, Xavier Cartoixà

Departament d'Enginyeria Electrònica, Escola d'Enginyeria, Universitat Autònoma de Barcelona, Campus UAB, E-08193 Bellaterra, Spain

## Abstract

Channel length scaling in graphene field effect transistors (GFETs) is key in the pursuit of higher performance in radio frequency electronics for both rigid and flexible substrates. Although two-dimensional (2D) materials provide a superior immunity to Short Channel Effects (SCEs) than bulk materials, they could dominate in scaled GFETs. In this work, we have developed a model that calculates electron and hole transport along the graphene channel in a drift-diffusion basis, while considering the 2D electrostatics. Our model obtains the self-consistent solution of the 2D Poisson's equation coupled to the current continuity equation, the latter embedding an appropriate model for drift velocity saturation. We have studied the role played by the electrostatics and the velocity saturation in GFETs with short channel lengths $L$. Severe scaling results in a high degradation of GFET output conductance. The extrinsic cutoff frequency follows a $1/L^n$ scaling trend, where the index $n$ fulfills $n \leq 2$. The case $n = 2$ corresponds to long-channel GFETs with low source/drain series resistance, that is, devices where the channel resistance is controlling the drain current. For high series resistance, $n$ decreases down to $n = 1$, and it degrades to values of $n < 1$ because of the SCEs, especially at high drain bias. The model predicts high maximum oscillation frequencies above 1 THz for channel lengths below 100 nm, but, in order to obtain these frequencies, it is very important to minimize the gate series resistance. The model shows very good agreement with experimental current voltage curves obtained from short channel GFETs and also reproduces negative differential resistance, which is due to a reduction of diffusion current.

# 1. Introduction

Graphene field-effects transistors (GFETs) are expected to become very relevant in radio frequency (RF) electronics [1–3] because of the exceptional intrinsic properties of the graphene: a carrier mobility over $10^5$ cm$^2$ V$^{-1}$ s$^{-1}$ and a saturation velocity of about $10^8$ cm s$^{-1}$ [4]. Theoretical studies have predicted GFETs to be able to work above the THz regime [5]. Relevant figures of merit (FoMs) in the field of high frequency electronics are the cutoff frequency $f_T$ and the maximum oscillation frequency $f_{max}$, the latter even more important than the former in practical applications. Reaching frequencies of 1 THz in these FoMs would mean large improvements in fields such as high speed communications or medicine [6]. GFETs have experimentally demonstrated a $f_T$ of 427 GHz for a 65-nm-long channel [7], which competes with other RF transistors. The record currently belongs to GaAs based high electron mobility transistors (HEMT), which have shown a $f_T$ of 688 GHz at a channel length of 40 nm [8]. Besides, GFET maximum $f_T$ has increased at a high pace during the last years. By contrast, $f_{max}$ has only reached 105 GHz for GFETs [9], while InP HEMTs have surpassed 1 THz [10]. In spite of the better RF performance of HEMTs and Si transistors, their application is limited to rigid electronics since the substrates can only withstand strains up to 1.08% [11]. Graphene presents high strain limits (up to 25%), so this two-dimensional (2D) material promises to stand out in the field of flexible electronics [12].

The upscaling of these FoMs of RF electronics implies a progressive shrinking of the channel length. Depending on the circuit application, the GFET is biased at the appropriate DC point that, for instance, maximizes the transconductance-output conductance ratio $g_m/g_d$ (to make a high voltage gain amplifier). The RF FoMs strongly depend on the particular values of $g_m$ and $g_d$. A lower value of $g_m$ will result in a lowering of $f_T$, while a higher value of $g_d$ will lower $f_{max}$. Short channels and/or high biases can make $g_m$ and $g_d$ strongly deviate from the long channel prediction, even if the 2D character of graphene provides a superior control over the charge in the channel when compared to bulk materials. This is due to the fact that SCEs degrade the ability of the gate to control electrostatically the carrier concentration in the channel. In general, SCEs worsen the FoMs and can have an important effect, as we will demonstrate along this work. This way, a model taking into account these effects becomes essential to make realistic predictions on the RF FoMs of scaled GFETs. Fabrication of short channel GFETs remains a major challenge because of the degradation of graphene mobility and a contact resistance comparable to channel resistance. Different approaches based on the self-alignment process can be applied to fabricate ultra-fast transistors, e. g. with a nanowire gated GFET [13], a pre-deposited gold protection layer on graphene [9] or transferring the whole gate stack [7].

Since the fabrication of the first GFET [14], multiple analytical models have been developed to describe its transistor effect [15–20]. These models give insight into the physics of graphene but they do not take account of the 2D electrostatics. In this work, we have developed a model that solves self-consistently the 2D Poisson's equation together with the current continuity equation in order to analyze short channel transistors using a drift-diffusion transport theory, which is valid as long as the transistor remains out the ballistic regime [21]. This means that the channel length must be larger than the mean free path of the carriers within graphene, estimated to be 10-100 nm [22–24]. The solution of the 2D potential equation allows for SCEs to be investigated, including the electrostatic influence of the drain in determining the carrier distribution along the channel. We have not considered impact ionization or hot carrier effects in the model. Impact ionization could result in an "up-kick" of the current-voltage characteristics, implying a worsening of $g_d$ [25]. In that case the FoMs predicted by our model should be regarded as an upper limit. For their part, hot carriers may arise in graphene from photoexcitation [26] or from a very thin insulator as in the graphene base transistor [27], but none of these characteristics are present in the GFETs we are considering. High electric fields could also provide high energy to carriers and this would imply a quasi-ballistic conductance that falls out of the scope of our model. Nevertheless, we find a good agreement between current curves from our numerical model (without impact ionization) and the experiment of a 70 nm short-channel GFET [28] (see section 3.5). Thus, we suspect that the strength of impact ionization and hot carriers depends not only on the critical field but on the quality of the graphene sample and substrate interaction effects. This model is also able to reproduce the negative differential resistance (NDR), observed experimentally in GFETs [29,30], which is a valuable property for many applications [29].

The paper is divided as follows: in Section 2, we explain the model and the way RF behavior is calculated. Section 3 is devoted to results. First, we analyze the electrostatics and current-voltage characteristics of short-channel GFETs in contrast with long-channel devices. Then, we explore the role played by the velocity saturation in the SCEs. Next, we discuss how the RF FoMs scale with the channel length and we comment on the differences with respect to long-channel models. To end up Section 3, we present a comparison with an experimental device and we study some devices showing NDR and analyze the physics behind it. Finally, Section 4 exposes the main conclusions of the work.

## 2. Methods

### 2.1 GFET model

As other models, our method of analyzing the electrical behavior of GFETs takes account of the basic physical principles of thermal statistics, electrostatics and current continuity [31,32]. However it adds the contribution of 2D electrostatics coupled to the one dimensional (1D) drift-diffusion transport equation.

Figure 1 (a) presents the physical structure of the double-gate GFET considered in this work. The graphene sheet plays the role of the active channel between the source and the drain. The device width ($W$) along the $z$-axis is large enough to consider that the transistor is uniform in that direction. To get the electrostatic behavior, Poisson's equation needs to be solved in a 2D region as the one depicted in figure 1 (b). Direction $x$ goes from the back gate to the top gate, while direction $y$ extends from source to drain along the channel length ($L$). Poisson's equation then takes the following form [33]:

$$\mathbf{\nabla} \cdot [\epsilon_r(x,y)\epsilon_0 \mathbf{\nabla}\psi(x,y)] = \rho_{\text{free}}(x,y) \tag{1}$$

where $\rho_{\text{free}}$ is the free charge density, $\epsilon_r$ is the relative dielectric permittivity of the medium and $\epsilon_0$ is the vacuum permittivity; $\psi$ is the electrostatic potential, which is directly related to the local position of the Dirac energy $E_D = -q\psi$, and $q$ is the elementary charge. In figure 1 (b), the thickness and the relative permittivity of the top oxide are $t_t$ and $\epsilon_{rt}$. The parameters $t_b$ and $\epsilon_{rb}$ correspond to those of the bottom oxide. The source is connected to ground and represents the reference potential, while the drain is connected to $V_{ds}$. Both $V_{gs}$ and $V_{bs}$ are the top and back gate voltages, respectively. Gate voltages electrostatically modulate the carrier concentration in graphene. The free charge $\rho_{\text{free}}(x,y)$ consists of the sheet charge density in graphene $\sigma(y) = q(p-n)$, where $n$ and $p$ are the electron and hole sheet densities respectively.

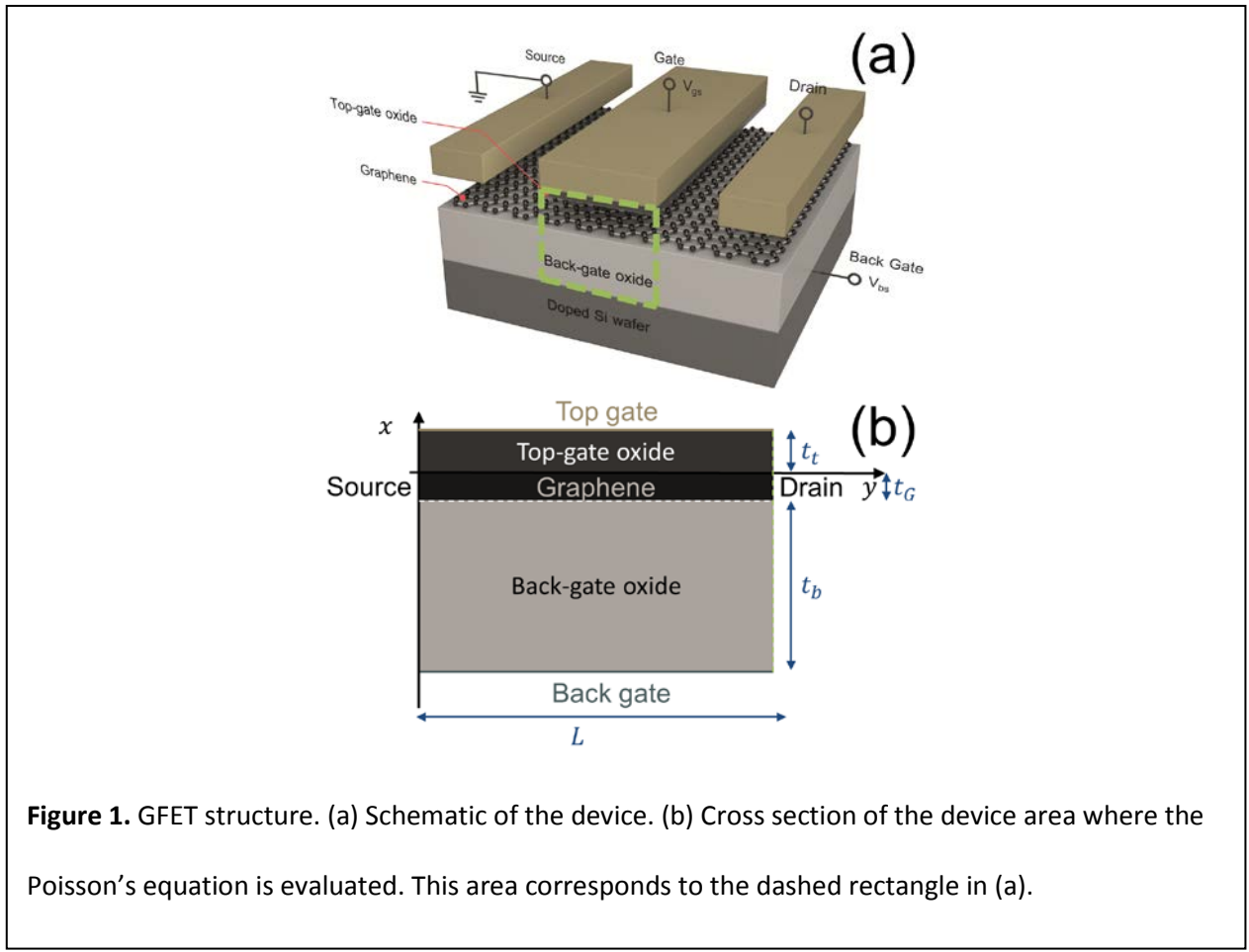

**Figure 1.** GFET structure. (a) Schematic of the device. (b) Cross section of the device area where the Poisson's equation is evaluated. This area corresponds to the dashed rectangle in (a).

To solve the 2D Poisson's equation, Dirichlet boundary conditions are applied at the gate contacts ($V_{gs} - V_{gs0}$ at the top gate and $V_{bs} - V_{bs0}$ at the back gate). Flatband voltages $V_{gs0}$ and $V_{bs0}$ are subtracted from $V_{gs}$ and $V_{bs}$, respectively, to take into account any possible fixed trapped charge at the oxides [31,34].

Von Neumann homogeneous boundary conditions are assumed at the rest of domain boundaries. At the boundaries of graphene with the source and drain, von Neumann conditions ensure charge neutrality since the electric potential profile can float at those points [35].

Our method includes the 1D expression of the drain current $I_{ds}$ in the diffusive regime, which results from the quasi-Fermi potential (or local Fermi energy) $E_F = -qV$ gradient [31]:

$$I_{ds} = qW(n\mu_n + p\mu_p)\frac{dV}{dy} \tag{2}$$

The values $\mu_n$ and $\mu_p$ correspond to electron and hole mobilities, respectively, and can vary along the channel. In equation (2), a single quasi-Fermi level has been considered for both electrons and holes. This is a valid approximation for graphene since generation/recombination times are very short (1-100 ps) [36–38] and thus electron and hole quasi Fermi levels cannot deviate too much from each other [31]. The boundary conditions state that $V(y)$ must be zero at the source ($y = 0$) and $V_{ds}$ at the drain ($y = L$). The difference between electrostatic and electrochemical potentials is called the chemical potential $V_c = \psi - V$, which is directly related to the carrier concentration inside graphene. To calculate electron and hole mobilities under high fields, a standard velocity saturation model has been considered under the form:

$$\mu_{n,p} = \frac{\mu_{0n,p}}{\left[1 + \left(\frac{\mu_{0n,p}}{v_{\text{sat}n,p}}\left|\frac{\partial \psi}{\partial y}\right|\right)^{\beta_{n,p}}\right]^{1/\beta_{n,p}}} \tag{3}$$

where $\mu_{0n}$ and $\mu_{0p}$ are the low-field mobilities for electrons and holes respectively and their values include the influence of the surface scattering or any other scattering mechanism, $v_{\text{sat},n}$ and $v_{\text{sat},p}$ are saturation carrier velocities, $-\partial\psi/\partial y$ is the electric field along the $y$ direction, and $\beta_n$ and $\beta_p$ are adimensional saturation coefficients for both kinds of carriers. This model was proposed for silicon transistors [39] and it has been successfully applied to graphene transistors [4,19]. It is important to note that $I_{ds}$ must fulfill the continuity condition, so, if we consider stationary situations, $I_{ds}$ must be constant along the channel ($dI_{ds}/dy = 0$).

Our method solves equations (1) and (2) in a self-consistent way starting from an initial guess of the graphene charge density $\sigma_0(y)$ (with $0 < y < L$). The algorithm is fully detailed in the appendix. It is able to find the drain current $I_{ds}$, the electrostatic and electrochemical potentials $\psi(x,y)$ and $V(y)$, and the carrier distributions, $n(y)$ for electrons and $p(y)$ for holes, for any given bias point.

## 2.3 High frequency performance calculation

GFET high frequency performance has been studied through the following FoMs: $f_T$, the frequency at which the current gain becomes unity; $f_{max}$, the frequency at which the power gain becomes unity; and the intrinsic voltage gain $A_v$. These three magnitudes can be extracted from a small signal model explained in ref. [40] and can be calculated from the following expressions:

$$f_{T,i} = \frac{|g_m|}{2\pi(C_{gs} + C_{gd})} \tag{4}$$

$$f_{T,x} = \frac{f_{T,i}}{1 + g_d(R_S + R_D) + \frac{C_{gd}g_m(R_S + R_D)}{C_{gs} + C_{gd}}} \tag{5}$$

$$f_{max} = \frac{|g_m|}{2\pi C_{gs}} \frac{1}{\sqrt{g_d(R_S + R_G + R_i) + \frac{C_{gd}R_G g_m}{C_{gs}}}} \tag{6}$$

$$A_v = \left|\frac{g_m}{g_d}\right| \tag{7}$$

As defined in ref. [40], the parameters used in equations (4)-(7) are the transconductance $g_m = \partial I_{ds}/\partial V_{gs}$, the drain conductance $g_d = \partial I_{ds}/\partial V_{ds}$, the gate-source capacitance $C_{gs} = \partial Q_{ch}/\partial V_{gs}$, and the gate-drain capacitance $C_{gd} = \partial Q_{ch}/\partial V_{ds}$. The intrinsic cutoff frequency $f_{T,i}$ just refers to the intrinsic device. On the other hand, the extrinsic cutoff frequency $f_{T,x}$ and $f_{max}$ depend on the series resistances at source and drain, $R_S$ and $R_D$. Moreover, $f_{max}$ depends on the series resistance at the gate $R_G$ and the internal resistance $R_i$. The latter has been neglected for the sake of simplicity. In order to obtain the capacitances $C_{gs}$ and $C_{gd}$, we must calculate the total charge in the channel $Q_{ch}$ for each bias point by means of the following equation:

$$Q_{ch} = W \int_0^L \sigma(y) dy \tag{8}$$

The aforementioned FoMs strongly depend on the bias conditions. In the supplementary material, we show the values of the parameters $g_m$, $g_d$, $C_{gs}$ and $C_{gd}$ for a reference device (figure S1) and we explain how to select the bias point to get these FoMs (figure S2).

## 3. Results and discussion

In order to test the model, we have simulated a GFET fabricated on a 400 nm thick SiO$_2$ (on a highly doped Si wafer that serves as back gate), with a silicon oxide of 40 nm as top gate insulator. Table 1 summarizes the reference GFET parameters used in this work. The temperature was taken as 300 K along this study. Section 3.5 benchmarks our model against experimental results of a GFET with a 70 nm long channel from ref. [28].

**Table 1.** GFET parameters used through this work. In case one parameter is varied, it is specified in the text.

| Parameter | Value |
|---|---|
| $t_t$ | 40 nm |
| $t_b$ | 400 nm |
| $\epsilon_{rt}$ | 3.9 |
| $\epsilon_{rb}$ | 3.9 |
| $V_{gs0}$ | 0.85 V |
| $V_{bs0}$ | 0 V |
| $\mu_{n0}, \mu_{p0}$ | 7500 cm$^2$ V$^{-1}$ s$^{-1}$ |
| $v_{sat,n}, v_{sat,p}$ | $v_F = 10^8$ cm s$^{-1}$ |

### 3.1 Short channel effects

Output and transfer characteristics ($I_{ds}$ - $V_{ds}$ and $I_{ds}$ - $V_{gs}$, respectively) calculated from our model are presented in figure 2 for channel lengths of 1 µm, 300 nm and 30 nm. Although the estimated carrier mean free path for the parameters used in this work is approximately 40 nm for a carrier concentration of $2 \cdot 10^{11}$ cm$^{-2}$ [22], we have simulated such short channels in order to enhance SCEs and observe them in a clearer way, assuming they are working in a diffusive regime. The results obtained can be regarded as a lower limit of the device performance. In order to check the validity of the method, figures 2 (a) and (d) benchmark the characteristics for the 1 µm device against the compact model reported in ref. [17]. Values slightly diverge, although they reproduce the main features of the $I_{ds}$ - $V_{ds}$ and $I_{ds}$ - $V_{gs}$ curves. The discrepancies come from the different approaches used to model both the quantum capacitance and the velocity saturation. The output characteristics show that drain current increases for shorter channel lengths, as expected. However, the increase rate is lower than $1/L$ because of both velocity saturation and 2D electrostatic effects. Moreover, $g_d$ increases, indicating poorer current saturation. The transfer characteristics show that $g_m$ improves with the channel length scaling except for very short channels (30 nm) at $V_{ds}$ over 0.5 V. In the $I_{ds}$ - $V_{gs}$ characteristics of figure 2 (c) the current loses its dependence on the gate voltage from that $V_{ds}$ onward. The behavior of $g_m$ and $g_d$ as a function of the channel length and the drain voltage is represented in the supplementary material (figure S3). We will see later that their degradation as $L$ scales down strongly affects the RF performance.

In ref. [41], S. Hen-Jan *et al.* found that, when the channel length is short, the Dirac voltage $V_D$ (the gate voltage that corresponds to the minimum current for a constant $V_{ds}$) shifted towards lower voltages when increasing $V_{ds}$. They related this to a weak gate control in short-channel devices. We expected to reproduce this result but Dirac voltages always shifts towards higher voltages with this model -see figure 2 (d), (e) and (f). It shows that the $V_D$ follows the rule $V_D = V_{gs0} + 1/2\, V_{ds}$ that is usually found [30].

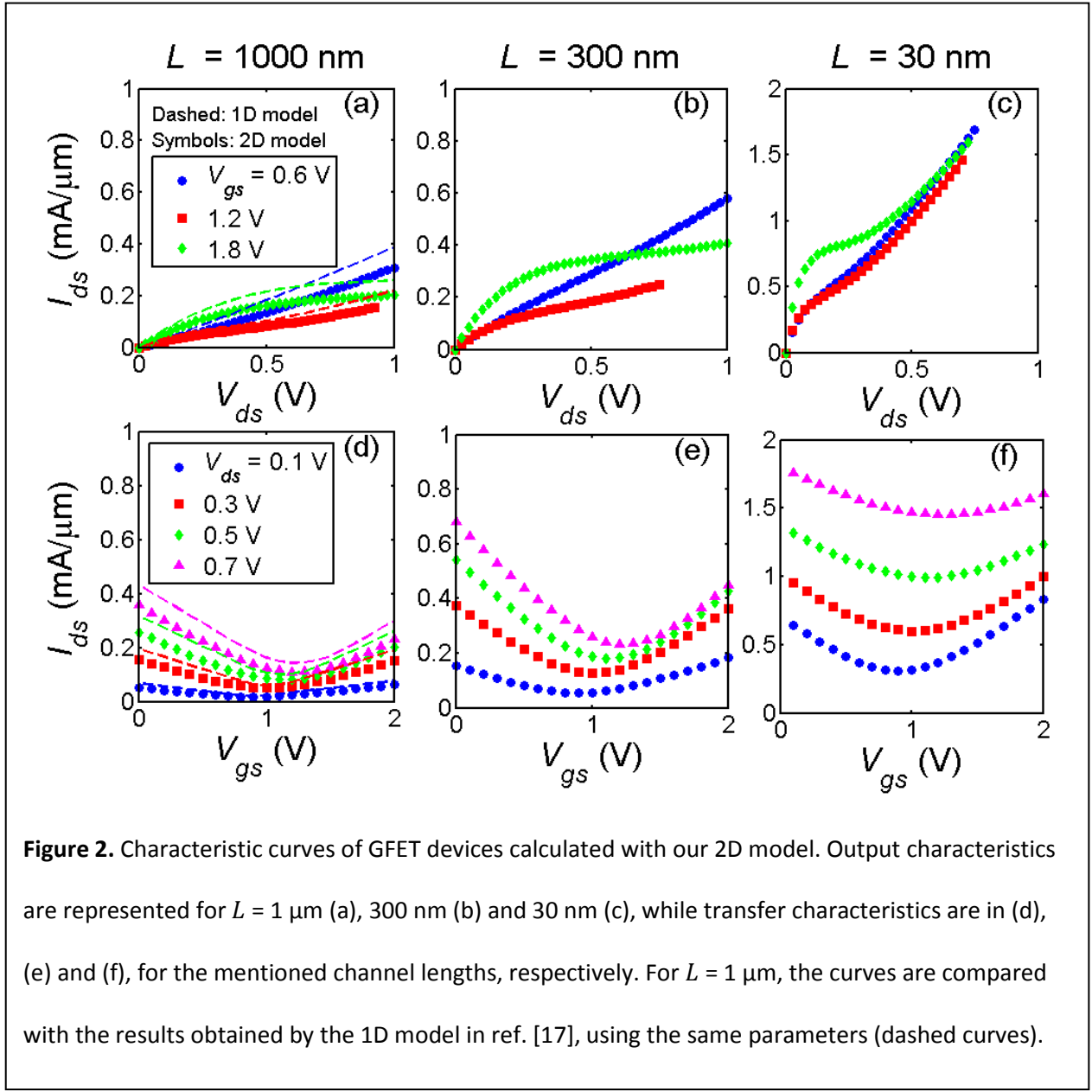

**Figure 2.** Characteristic curves of GFET devices calculated with our 2D model. Output characteristics are represented for $L$ = 1 μm (a), 300 nm (b) and 30 nm (c), while transfer characteristics are in (d), (e) and (f), for the mentioned channel lengths, respectively. For $L$ = 1 μm, the curves are compared with the results obtained by the 1D model in ref. [17], using the same parameters (dashed curves).

To get some insight into the physics of the GFET, we have represented in figure 3 the chemical potential and energy profiles along the channel for different lengths at several bias points. In all cases, the top-gate voltage has been assumed to be 1.2 V. For the long-channel GFETs at low drain bias no pinch-off occurs in the channel, that is, all carriers are of the same type (electrons in this case) and thus $V_c$ does not change its sign. As the drain voltage increases, holes start to appear close to the drain. When the drain voltage is 0.6 V, the pinch-off point (where $V_c$ = 0) is approximately in the middle of the channel. For very short channels (30 nm), a drain voltage as low as 0.2 V is enough to cause pinched-off in the channel. In this case, carrier concentrations close to the source and drain are also much higher. Due to the short channel, the drain voltage affects the electrostatic potential within the entire device, even in the region close to the source. So the SCEs in GFETs manifest in the following ways: (1) an increase of the carrier concentration close to the source and drain, (2) a more abrupt chemical potential slope along the channel and (3) a shift of the pinch-off point inside the graphene towards the middle of the channel.

In figure 4, the electrostatic potential distribution is represented for several channel lengths. All three devices show the pinch-off point around the middle of the channel. The potential profile varies as channel length scales down: the isopotential lines (black lines) accumulate close to the graphene for shorter channel lengths. The higher electric field in the graphene leads to the higher carrier concentrations and the more abrupt transitions of $V_c$ around the pinch-off.

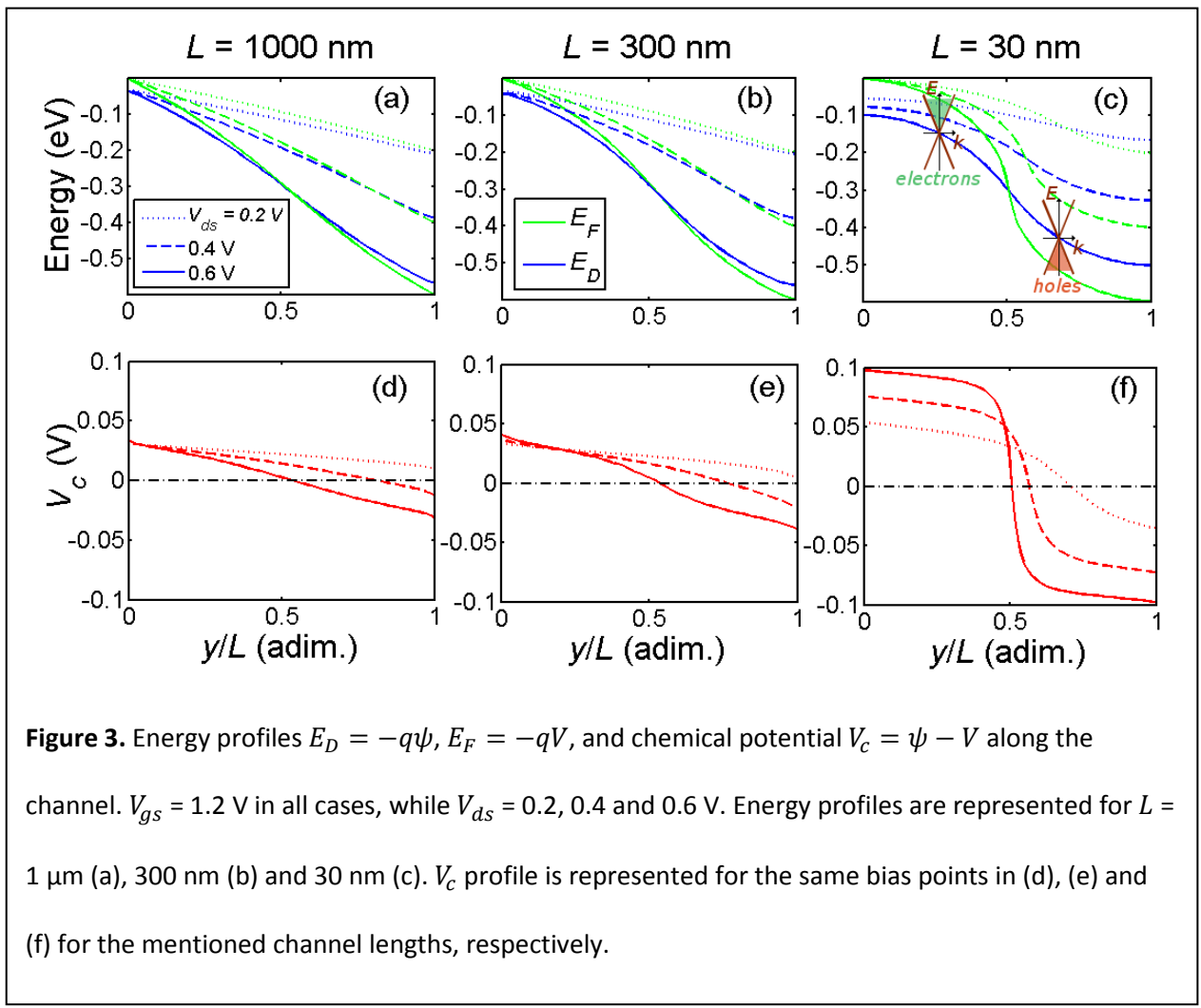

**Figure 3.** Energy profiles $E_D = -q\psi$, $E_F = -qV$, and chemical potential $V_c = \psi - V$ along the channel. $V_{gs}$ = 1.2 V in all cases, while $V_{ds}$ = 0.2, 0.4 and 0.6 V. Energy profiles are represented for $L$ = 1 μm (a), 300 nm (b) and 30 nm (c). $V_c$ profile is represented for the same bias points in (d), (e) and (f) for the mentioned channel lengths, respectively.

To end this subsection, it is worth noting that punch-through effect is not observable in GFETs. In Si based devices, it occurs when source and drain depletion regions touch each other and causes a total loss of gate modulation of drain current [42]. Graphene channels, by contrast, present no depletion regions.

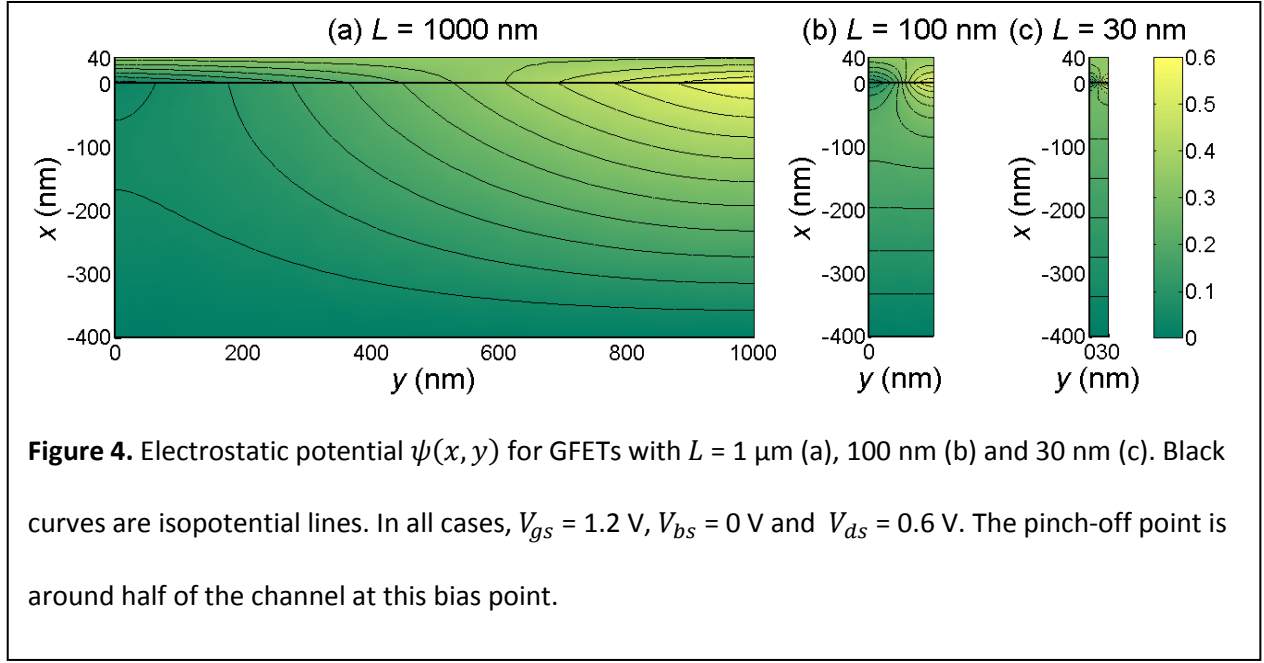

**Figure 4.** Electrostatic potential $\psi(x,y)$ for GFETs with $L$ = 1 μm (a), 100 nm (b) and 30 nm (c). Black curves are isopotential lines. In all cases, $V_{gs}$ = 1.2 V, $V_{bs}$ = 0 V and $V_{ds}$ = 0.6 V. The pinch-off point is around half of the channel at this bias point.

### 3.2 Impact of top oxide thickness and permittivity

Figure 5 summarizes the influence of the top gate oxide on the $I_{ds}$ - $V_{gs}$ and $I_{ds}$ - $V_{ds}$ characteristics of a 100 nm device. In all graphs, the filled symbols represent the reference device. This device is compared to others with different top oxide thickness, relative dielectric permittivity and equivalent oxide thickness (EOT). The EOT is equal to the SiO$_2$ thickness necessary to have the same gate oxide capacitance. In figures 5 (a) and (d), the oxide thickness is increased from 40 to ~90 nm. The transfer characteristics in figure 5 (a) show that the transconductance decreases while the current at the Dirac voltage remains almost the same. This means that a thicker oxide worsens gate control, which degrades RF performance and on/off ratio. Poor saturation is also observed in the $I_{ds}$ - $V_{ds}$ characteristics of figure 5 (d). A narrower top gate oxide is then desirable to improve RF performance.

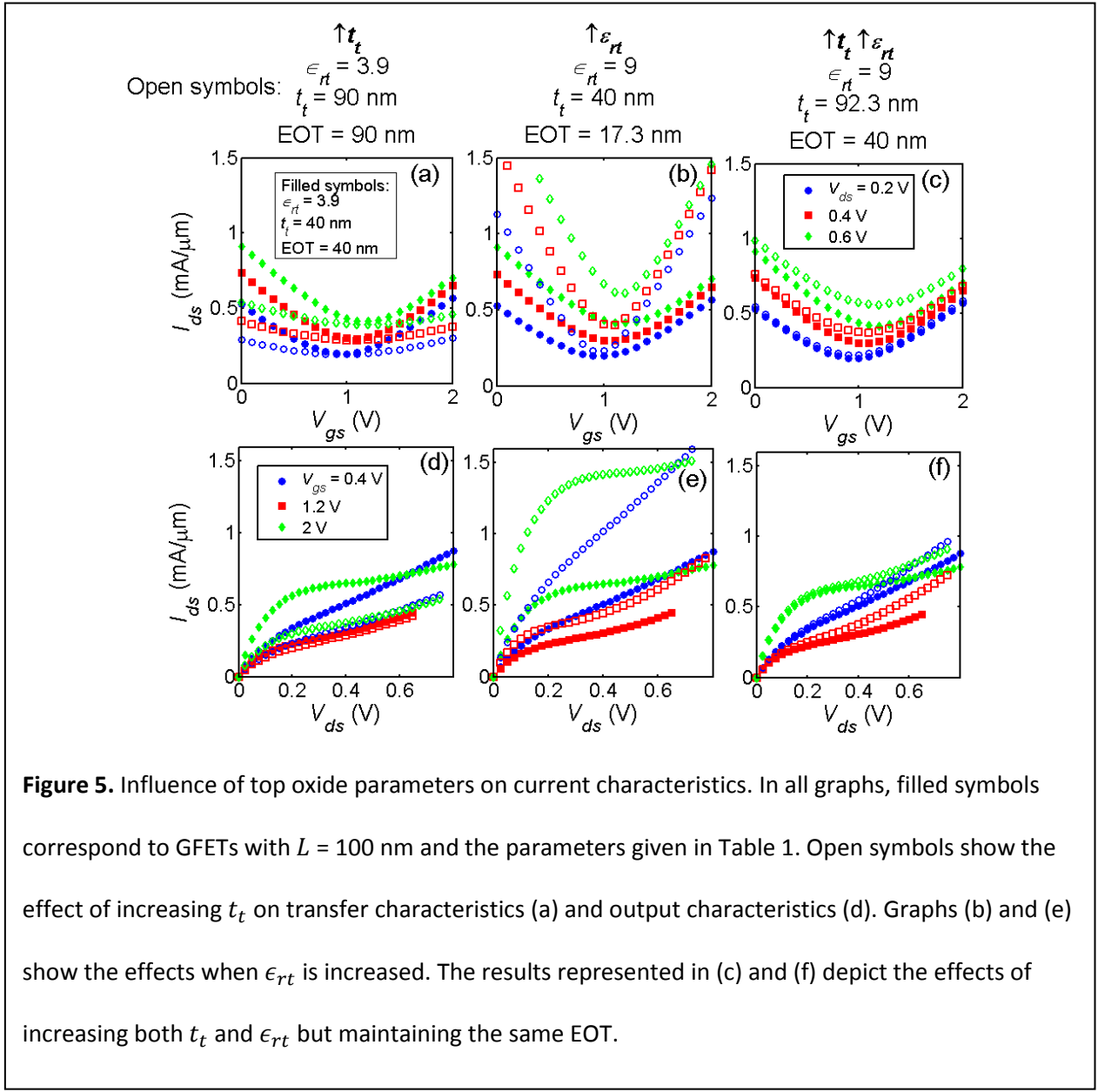

**Figure 5.** Influence of top oxide parameters on current characteristics. In all graphs, filled symbols correspond to GFETs with $L$ = 100 nm and the parameters given in Table 1. Open symbols show the effect of increasing $t_t$ on transfer characteristics (a) and output characteristics (d). Graphs (b) and (e) show the effects when $\epsilon_{rt}$ is increased. The results represented in (c) and (f) depict the effects of increasing both $t_t$ and $\epsilon_{rt}$ but maintaining the same EOT.

Figures 5 (b) and (e) compare the reference device with another where the relative dielectric constant of the top oxide is increased from 3.9 to 9. This means a reduction in the EOT from 40 to ≈ 17 nm. The $I_{ds}$ - $V_{gs}$ curves indicate that the transconductance improves with this change, although the minimum current increases. $I_{ds}$ - $V_{ds}$ curves in Figure 5 (e) show that increasing the dielectric permittivity leads to a worse current saturation. Figures 5 (c) and (f) display the characteristics of a device where both the top oxide thickness and its relative permittivity are increased, keeping the same EOT. We would expect that this did not affect the $I_{ds}$ - $V_{gs}$ and $I_{ds}$ - $V_{ds}$ curves, as it happens with long-channel Si-based devices [42]. In the supplementary material (figure S4) it is shown that this is the same case for long-channel GFETs, so this dependence shown in figures 5 (c) and (f) has to do with SCEs.

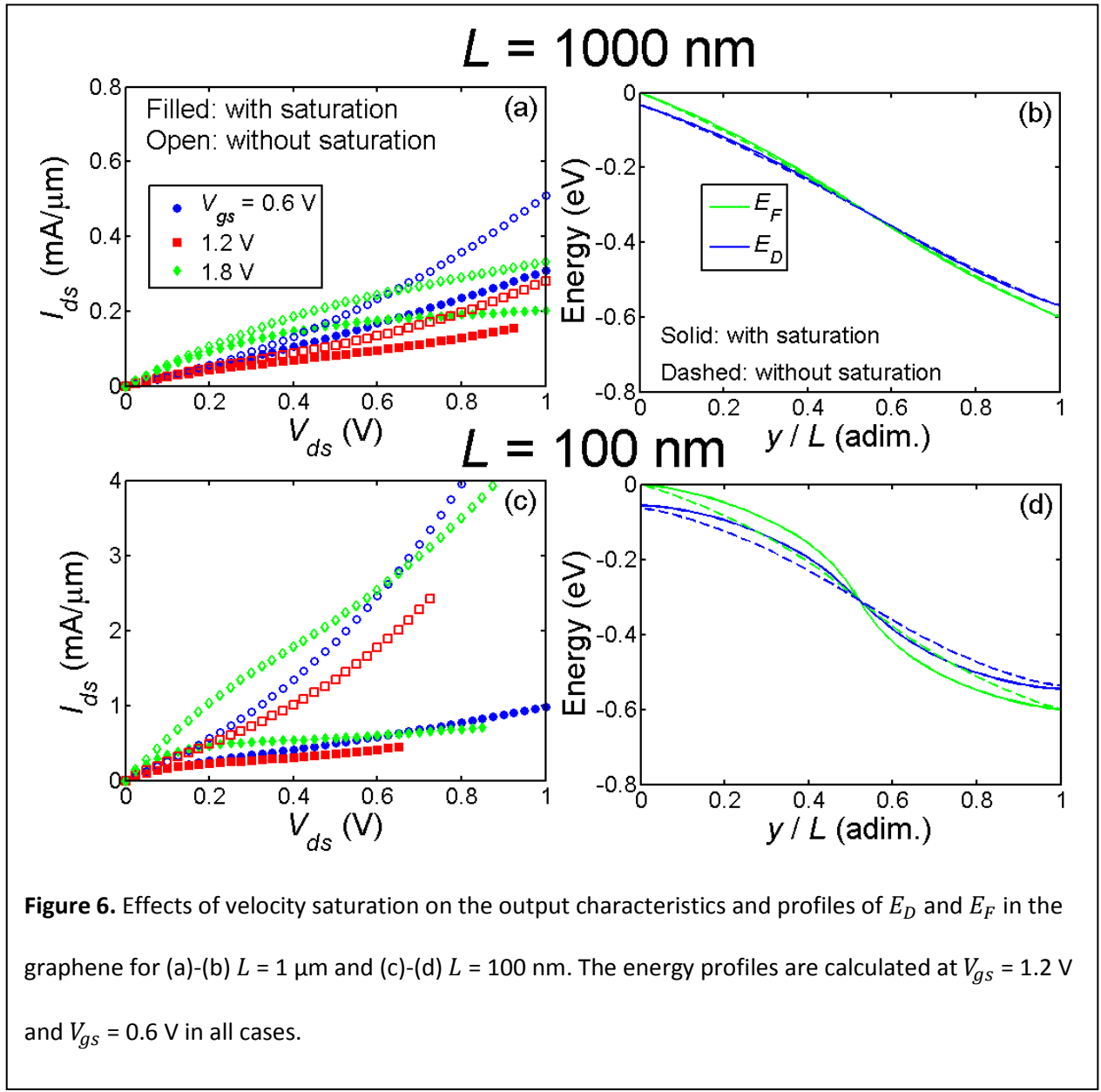

**Figure 6.** Effects of velocity saturation on the output characteristics and profiles of $E_D$ and $E_F$ in the graphene for (a)-(b) $L$ = 1 μm and (c)-(d) $L$ = 100 nm. The energy profiles are calculated at $V_{gs}$ = 1.2 V and $V_{gs}$ = 0.6 V in all cases.

### 3.3 Effect of carrier velocity saturation

In order to assess the influence of the velocity saturation, we have simulated devices assuming a constant low-field carrier mobility in equation (2). Figure 6 shows the $I_{ds}$ - $V_{ds}$ characteristics and energy profiles at different gate voltages for channel lengths of 1 μm and 100 nm, comparing the results with those that include velocity saturation. Energy profiles coincide for long channel regardless of saturation, as shown in figure 6 (b). However, the velocity saturation clearly influences the energy profiles for short lengths. It results in a different carrier distribution within the channel that affects the $I_{ds}$ - $V_{ds}$ characteristics. In fact, the only effect of velocity saturation in long-channel devices is a reduction in $I_{ds}$ of a factor $\gamma = 1 + \mu_{n0}V_{ds}/(v_F L)$, which can be interpreted as an effective channel length $L_{\text{eff}} = \gamma L$. By contrast, in short-channel GFETs, there is an additional loss of current not explained just by this $L_{\text{eff}}$. Velocity saturation also modifies the carrier distribution along the channel, and should be taken into account for accurate calculations.

### 3.4 High frequency performance

In this sub-section, we deal with the impact that SCEs have in the RF behavior. Specifically, figure 7 depicts $f_{T,i}$, $f_{T,x}$, $f_{\max}$ and $A_v$ as a function of the channel length. We reach lengths as low as 10 nm, but always assuming that the device is working in a diffusive regime. The understanding of drift-diffusion at short channels could help in a further study of the transition to the ballistic regime [43]. In figure 7 (a), $f_{T,i}$ is represented for two $V_{ds}$ (0.1 and 0.6 V). The dashed black lines represent $f_{T,v_F}$, which is the physical limit that no cutoff frequency could surpass for GFETs [20,31]. It comes out from the supposition that maximum group velocity achievable in graphene is the Fermi velocity $v_F$, following the trend $f_{T,v_F} \approx v_F/(2\pi L)$. Dotted lines correspond to the extrapolation to lower channel lengths of the $f_{T,i}$ calculated according to the 1D model from ref. [17]. It scales as $\sim 1/L^n$ with $n = 2$ since the transconductance is proportional to $1/L$ while the gate capacitance ($C_g = C_{gs} + C_{gd}$) is proportional to $L$. This trend is followed by $f_{T,i}$ for channel lengths down to around 1 µm. Both $g_m$ and $C_g$ follow the expected trends for long channels (see figure S3 in the supplementary material). It is worth noting that the oxide capacitance dominates over the quantum capacitance in this device.

In short channel MOSFETs $f_{T,i}$ scales as $1/L$ because of channel velocity saturation, making the transconductance independent of $L$. In the GFETs we have studied, velocity saturation is not complete, causing $g_m \propto 1/L^n$ with $0 < n < 1$, as shown in figure S3 of the supplementary material, and therefore $f_{T,i} \propto 1/L^n$ with $1 < n < 2$. As the drain bias is reduced, the field in the channel decreases and the carriers are driven farther from velocity saturation, thus having a lower deviation from the $1/L^2$ behavior -see the 0.1 V vs the 0.6V curve in figure 7(a). Therefore, a low drain bias helps to reduce this deviation, so higher cutoff frequencies can be reached.

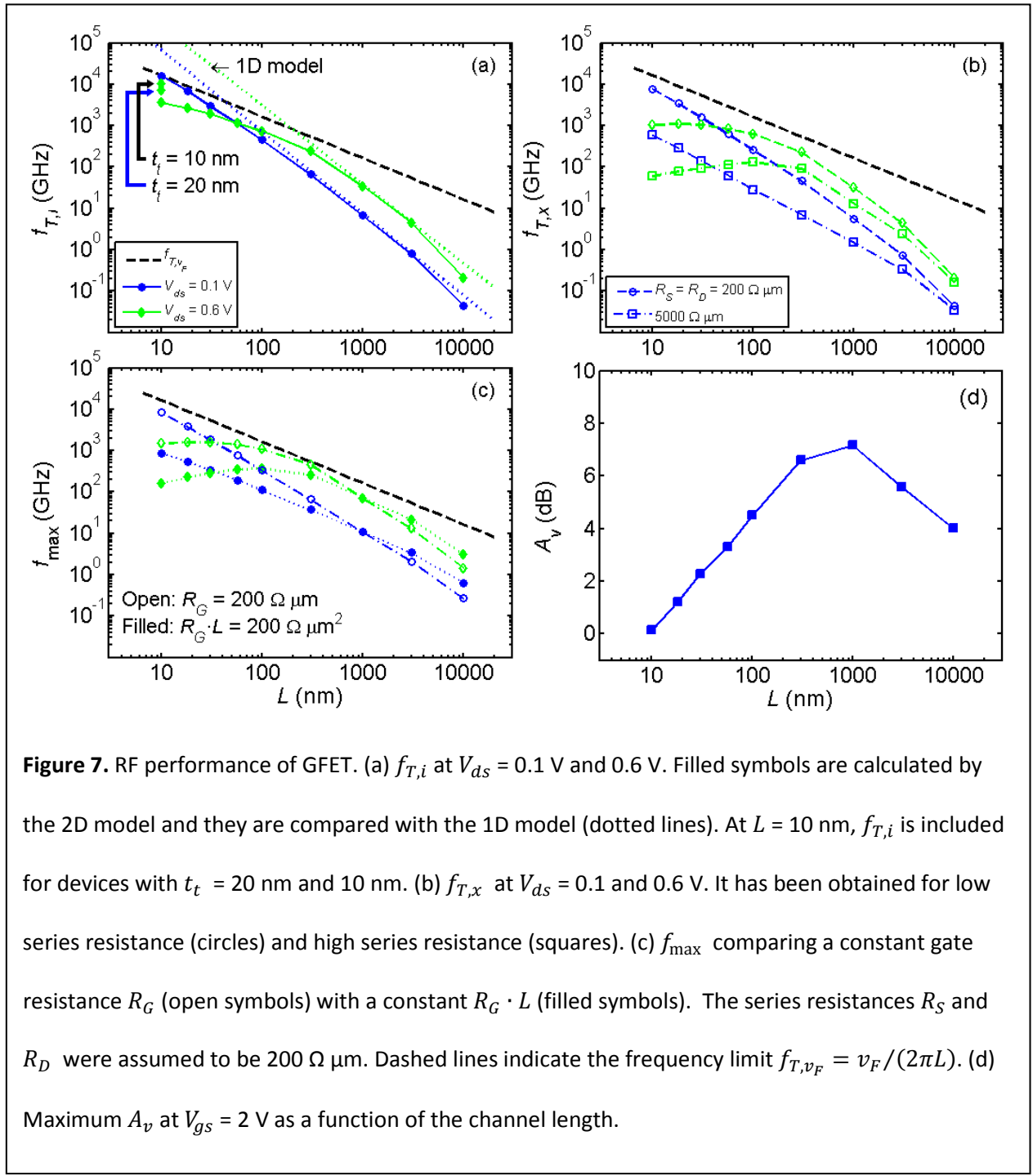

**Figure 7.** RF performance of GFET. (a) $f_{T,i}$ at $V_{ds}$ = 0.1 V and 0.6 V. Filled symbols are calculated by the 2D model and they are compared with the 1D model (dotted lines). At $L$ = 10 nm, $f_{T,i}$ is included for devices with $t_t$ = 20 nm and 10 nm. (b) $f_{T,x}$ at $V_{ds}$ = 0.1 and 0.6 V. It has been obtained for low series resistance (circles) and high series resistance (squares). (c) $f_{\max}$ comparing a constant gate resistance $R_G$ (open symbols) with a constant $R_G \cdot L$ (filled symbols). The series resistances $R_S$ and $R_D$ were assumed to be 200 Ω μm. Dashed lines indicate the frequency limit $f_{T,v_F} = v_F/(2\pi L)$. (d) Maximum $A_v$ at $V_{gs}$ = 2 V as a function of the channel length.

SCEs can be reduced at short channel lengths by an appropriate choice of $t_t$: figure 7 (a) shows that, for $L$ = 10 nm, a reduction in the top insulator thickness can help improve the cutoff frequency. Decreasing $t_t$ from 40 nm to 20 nm (or even 10 nm) does not increase $f_{T,i}$ at $V_{ds}$ = 0.1 V since it is close to the physical limit. However it improves from 3.6 THz to 10 THz at $V_{ds}$ = 0.6 V. Thus, a thin insulator is important to get the highest possible cutoff frequency. Similarly to silicon oxide in silicon-based devices, the lower limit of the insulator thickness will be set by the tolerable leakage current, which would affect both the power consumption and likely device reliability.

The extrinsic cutoff frequencies have been calculated assuming $R_S$ and $R_D$ of 200 Ω µm, which is currently a typical value of metal-graphene contact [44], and a higher value of 5000 Ω µm. When series resistance is low, its effects start to be important, as compared to $f_{T,i}$, for channel lengths below 100 nm. However, the case $R_S = R_D = 5000$ Ω µm implies that the current is dominated by the series resistance instead of the channel resistance. This reduces $n$, reaching values near 1 for low drain bias. Interestingly, Yin *et al.* found a cutoff frequency that scales with $n$ = 2 [45], which was attributed to a carrier transport limited by the channel resistance [46,47]. Later, the same group found a behavior of $n$ = 1 [48,49], which was attributed to a current limited by the source/drain contact resistance [47]. A brief summary of the $f_T$ scaling discussion can be found in Table 2.

**Table 2.** GFET cutoff frequency scaling.

|  | $f_{T,i} \propto 1/L^n$ | | $f_{T,x} \propto 1/L^n$ | | | |
|---|---|---|---|---|---|---|
|  | | | low $R_S, R_D$ | | high $R_S, R_D$ | |
|  | low $V_{ds}$ | high $V_{ds}$ | low $V_{ds}$ | high $V_{ds}$ | low $V_{ds}$ | high $V_{ds}$ |
| Long channel | $n \cong 2$ | $n \cong 2$ | $n \cong 2$ | $n \cong 2$ | $n \cong 1$ | $1 < n < 2$ |
| Short channel | $1 < n < 2$ | $n \lesssim 1$ | $1 < n < 2$ | $n < 1$ | $n \cong 1$ | $n < 1$ |

Regarding $f_{max}$, this magnitude is represented for the reference GFET in figure 7 (c). The loss of transconductance combined with a poorer current saturation in short channel devices limits the highest frequencies that can be achieved. It turns out that a low drain bias is helpful to get the highest maximum oscillation frequency. Two situations are compared in figure 7 (c): a constant $R_G$ (200 Ω µm) and a scaled $R_G$ ($R_G \cdot L$ = 200 Ω µm²). This scaled gate series resistance shows much lower $f_{max}$ at short channels and illustrates the importance of minimizing $R_G$ [20], which can be done in practice by adopting a T-gate architecture [50,51]. We find that $f_{max}$ predicted by our model can go beyond 1 THz for channel lengths shorter than 50 nm.

Figure 7 (d) shows maximum $A_v$ as a function of the channel length for the reference device. A maximum of around 7 dB can be found for a 1 µm GFET and then it degrades as $L$ becomes shorter. Such a degradation with $L$ has been experimentally found in previous works [52] and is mainly due to the loss of current saturation when the channel is made short.

## 3.5 Model comparison to experimental values

We have benchmarked the model of this work against an experimental 70 nm long GFET from ref. [28]. To simulate the device, we have used the parameters extracted experimentally. They can be found in Table 3. We have supposed a puddle concentration of $1.25 \cdot 10^{12}$ cm$^{-2}$, that is, a minimum carrier density, and a constant temperature of 300 K was taken. In figure 8 we represent both the simulation and the experimental curves. Specifically, figure 8 (a) represents the output characteristics of both the model and experiment, while figures 8 (b) and (c) compare transconductance and output conductance, respectively. The simulations and experiment show quantitative agreement even assuming in the simulation that the device is working in the drift-diffusion regime. We believe that the quality of the sample and/or interaction effects with the substrate is preventing the device to work in the ballistic regime.

**Table 3.** GFET parameters used to benchmark the model against experimental results [28].

| Parameter | Value |
|---|---|
| $L$ | 70 nm |
| $t_t$ | 400 nm |
| $t_b$ | 90 nm |
| $\epsilon_{rt}$ | 1 |
| $\epsilon_{rb}$ | 3.9 |
| $V_{gs0}$ | 15 V |
| $V_{bs0}$ | 0 V |
| $\mu_{n0}$ | 1250 cm$^2$ V$^{-1}$ s$^{-1}$ |
| $\mu_{p0}$ | 750 cm$^2$ V$^{-1}$ s$^{-1}$ |
| $v_{\text{sat},n}$, $v_{\text{sat},p}$ | $v_F = 10^8$ cm s$^{-1}$ |
| $\beta_n$, $\beta_p$ | 1 |
| $R_S$, $R_D$ | 280 Ω µm |

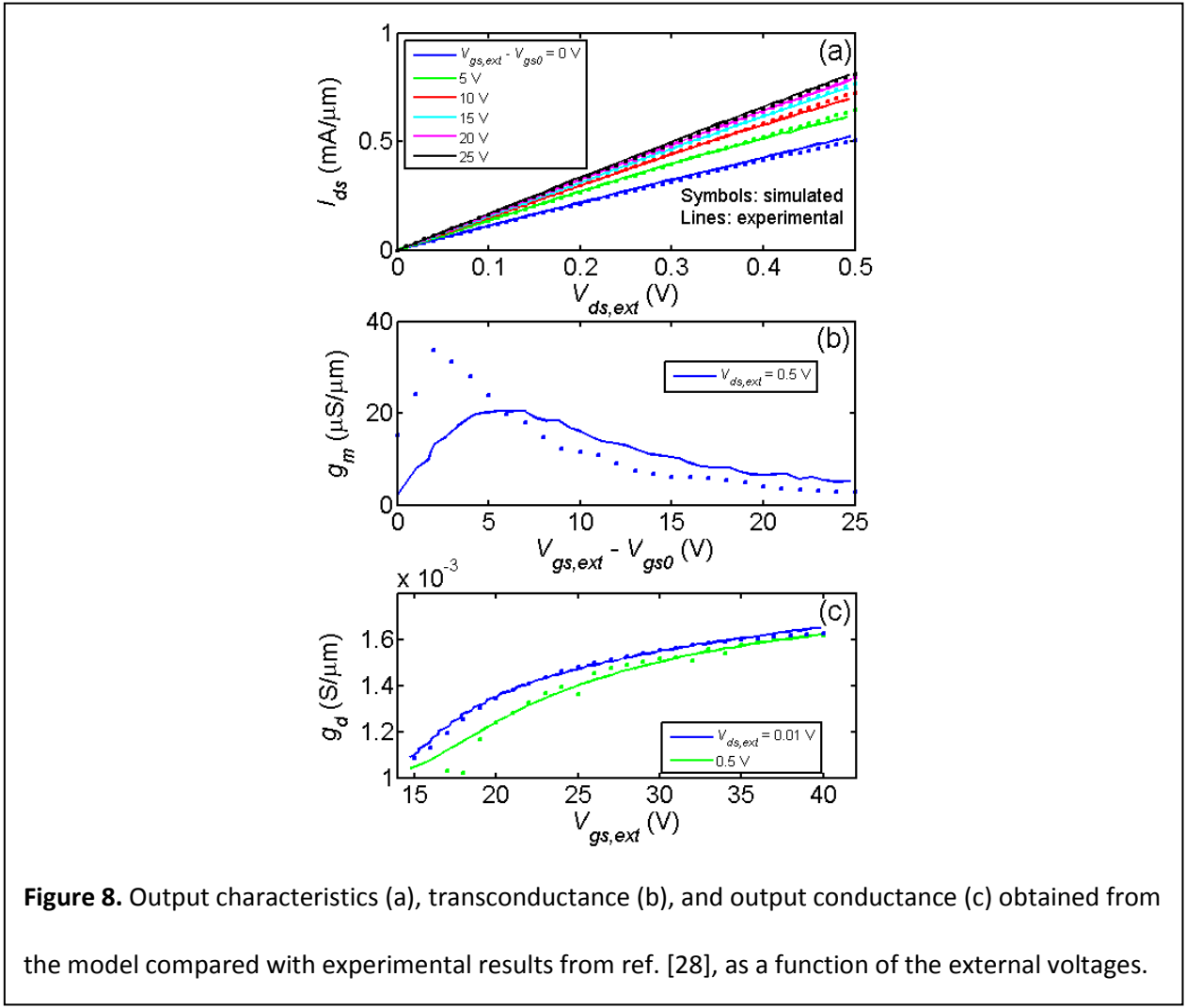

**Figure 8.** Output characteristics (a), transconductance (b), and output conductance (c) obtained from the model compared with experimental results from ref. [28], as a function of the external voltages.

### 3.6 Negative differential resistance

We have found that the model can describe NDR behavior, which has been experimentally observed in GFETs [29,30]. Figure 9 (a) represents the output characteristics for a device with a 100 nm channel length, a top-gate oxide thickness of 5.4 nm and a relative permittivity of 21. It compares two situations: $\beta = 1$ (open symbols) and $\beta = 2$ (filled symbols), with $\beta_n = \beta_p = \beta$. Although the former case shows NDR for gate voltages over 1.2 V, this effect is boosted when $\beta = 2$, where the maximum drift velocity is gotten at lower electric fields. Figure 9 (b) displays a single $I_{ds}$ - $V_{ds}$ curve where NDR arises. The current has been split into drift and diffusion contributions for both electrons and holes. At low $V_{ds}$, electron diffusion and drift currents increase until they reach a maximum. This happens while all carriers within the channel are electrons. As $V_{ds}$ further grows, the pinch-off approaches the drain contact. The drift component remains constant while the diffusion contribution starts to decrease. The hole contributions are still negligible, so this trend is the origin of the NDR. Then, the hole contribution starts to grow when the pinch-off moves away from the drain towards the source. The NDR effect is then caused because of the electron diffusion current reduction when the pinch-off point is around the drain. We have calculated the bias regime where the NDR comes out –shown in figure 9 (c). Because of SCEs, some differences arise with respect to the simple 1D model used in ref. [30].

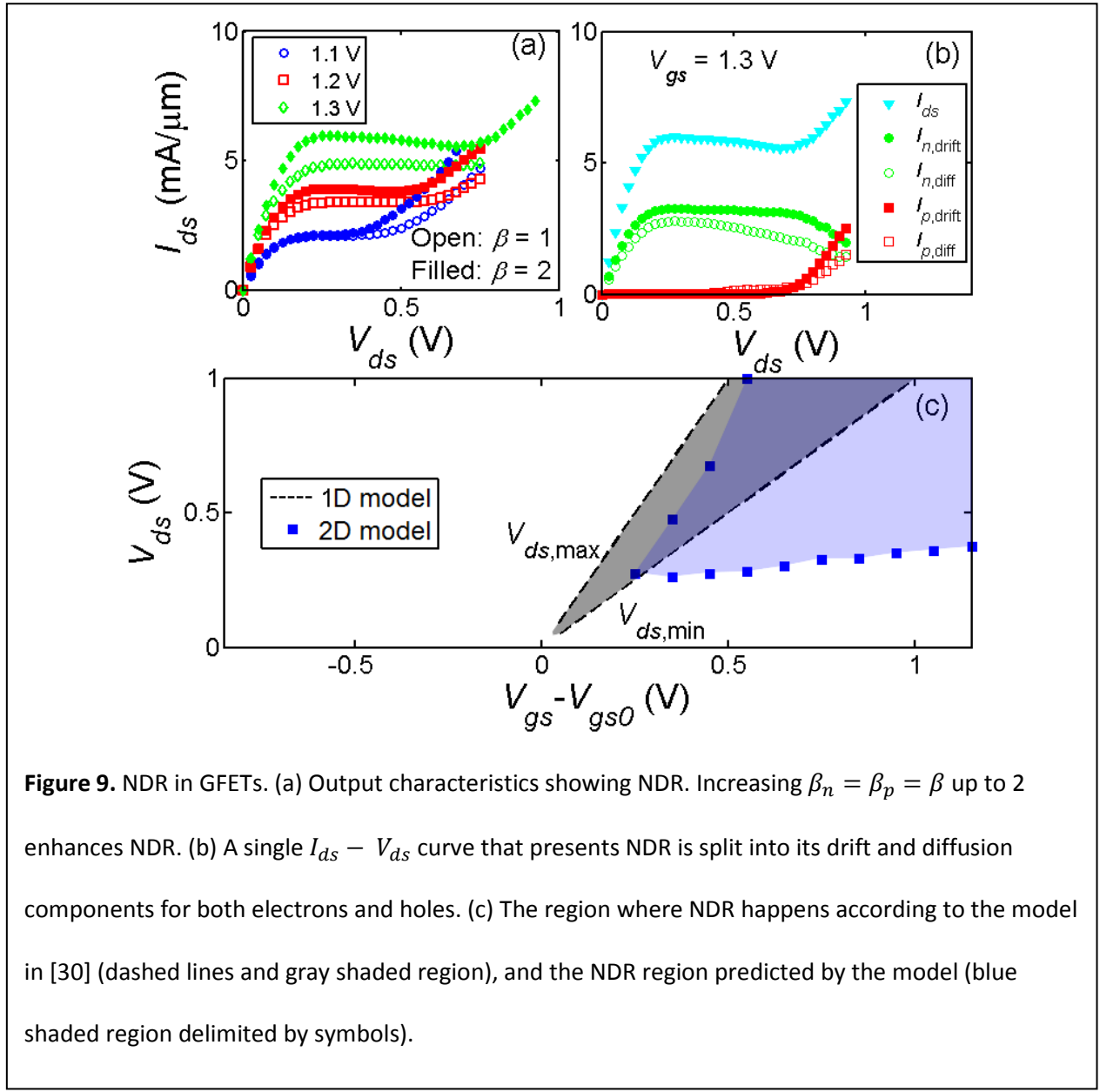

**Figure 9.** NDR in GFETs. (a) Output characteristics showing NDR. Increasing $\beta_n = \beta_p = \beta$ up to 2 enhances NDR. (b) A single $I_{ds} - V_{ds}$ curve that presents NDR is split into its drift and diffusion components for both electrons and holes. (c) The region where NDR happens according to the model in [30] (dashed lines and gray shaded region), and the NDR region predicted by the model (blue shaded region delimited by symbols).

## 4. Conclusions

In this work, we have developed a model that self-consistently solves the 2D Poisson's equation and the 1D drift-diffusion transport equation to investigate both the stationary and frequency response of GFETs. With this model we have studied SCEs, unveiling the role played by both the electrostatics and the velocity saturation effect. We have found that a degradation in both the transconductance and the output conductance in short-channel devices strongly affect the transistor RF behavior. We have found an extrinsic cutoff frequency scaling law that can be locally expressed as $1/L^n$ with $1 < n \leq 2$, where the scaling index has the mixed signature of SCEs and series resistance. The case $n$ = 2 corresponds to long-channel GFET with low series resistance, while $n$ = 1 takes place when the current is dominated by the series resistance. At high applied drain bias and short channels, the scaling index could even be $n < 1$, which severely limits the HF performance. Assuming state-of-the-art values for the source/drain and gate series resistances, we have found that cutoff and maximum oscillation frequencies above 1 THz needed a channel length well below 100 nm for our reference device. We have also studied the influence by the gate series resistance on the $f_{max}$ optimization and we have demonstrated that the impact of SCEs could be reduced by an appropriate choice of the top oxide thickness. Finally, it has been checked that the model compares very well to short channel experimental results and that it is able to capture NDR behavior and we have explained its origin.

## Acknowledgements

This project has received funding from the European Union's Horizon 2020 research and innovation programme under grant agreement No 696656, the *Department d'Universitats, Recerca i Societat de la Informació* of the *Generalitat de Catalunya* under contract 2014 SGR 384 and the *Ministerio de Economía y Competitividad* of Spain under grants TEC2012-31330 and TEC2015-67462-C2-1-R (MINECO/FEDER).

# Appendix. Self-consistent algorithm

We start the explanation of our model by recalling the principles on which it is based. According to the thermal statistics of graphene, the sheet concentrations $n$ and $p$ can be calculated by the following equations [31,32]:

$$n = N_G \mathcal{F}_1 \left[ \frac{E_F - E_D}{k_B T} \right] \tag{A.1}$$

$$p = N_G \mathcal{F}_1 \left[ \frac{E_D - E_F}{k_B T} \right] \tag{A.2}$$

In equations (A.1) and (A.2), $q$ is the elementary charge, $k_B$ is Boltzmann's constant, $T$ is the absolute temperature of the device, $N_G$ is the effective graphene sheet density of states, given by equation (A.3), and $\mathcal{F}_i(x)$ is the complete Fermi–Dirac integral with index $i$, given by equation (A.4).

$$N_G = \frac{2}{\pi} \left( \frac{k_B T}{\hbar v_F} \right)^2 \tag{A.3}$$

$$\mathcal{F}_i(x) = \frac{1}{\Gamma(i+1)} \int_0^\infty \frac{u^i du}{1 + e^{u-x}} \tag{A.4}$$

Here, $\hbar$ is the reduced Planck's constant, $v_F$ is the Fermi velocity of carriers in graphene ($10^8$ cm/s) and $\Gamma(i)$ is the gamma function. Equations (A.1) and (A.2) are deduced from the linear dispersion relation of graphene and the Fermi-Dirac thermal distribution [31,53].

In the Poisson's equation, equation (1), we have considered that the graphene has a thickness of $t_G$ = 0.6 nm and a relative dielectric permittivity of $\epsilon_{rG}$ = 3.3 [54]. The free charge $\rho_{\text{free}}(x, y)$ is then:

$$\rho_{\text{free}}(x, y) = \begin{cases} \dfrac{\sigma(y)}{t_G} & -t_G < x < 0 \text{ and } 0 < y < L \\ 0 & \text{inside the dielectrics} \end{cases} \tag{A.5}$$

Our method follows the steps as presented in the flow diagram of figure A.1. From an initial guess $\sigma_0(y)$ (with $0 < y < L$), the system is solved for a bias point (fixed $V_{gs}$, $V_{bs}$ and $V_{ds}$). The initial guess is obtained from previous 1D long channel models [17] or from a solution of the same device at the same gate voltage but at a lower $V_{ds}$.

Charge distribution $\sigma_k(y)$ at the iteration $k$ is used in Poisson's equation to determine the electric potential profile $\psi_k(x,y)$ within the whole domain. Equation 1 is solved inside the 2D region of figure 1 (b) by the advanced Finite Element Method of the Matlab® partial differential equation solver.

Then, the electrostatic profile $\psi_k(y) = \psi_k(0, y)$ is introduced into the drift-diffusion equation, equation (2), to determine the quasi-Fermi level profile $V_k(y)$ together with the drain current $I_{ds,k}$, which acts as a parameter in the first-order Ordinary Differential Equation (ODE). The integration of this equation is carried out by a first-order predictor-corrector method [55]. We use the shooting method to ensure that the quasi-Fermi potential reaches $V_{ds}$ at $y = L$, using an adapted bisection algorithm to find the parameter $I_{ds,k}$.

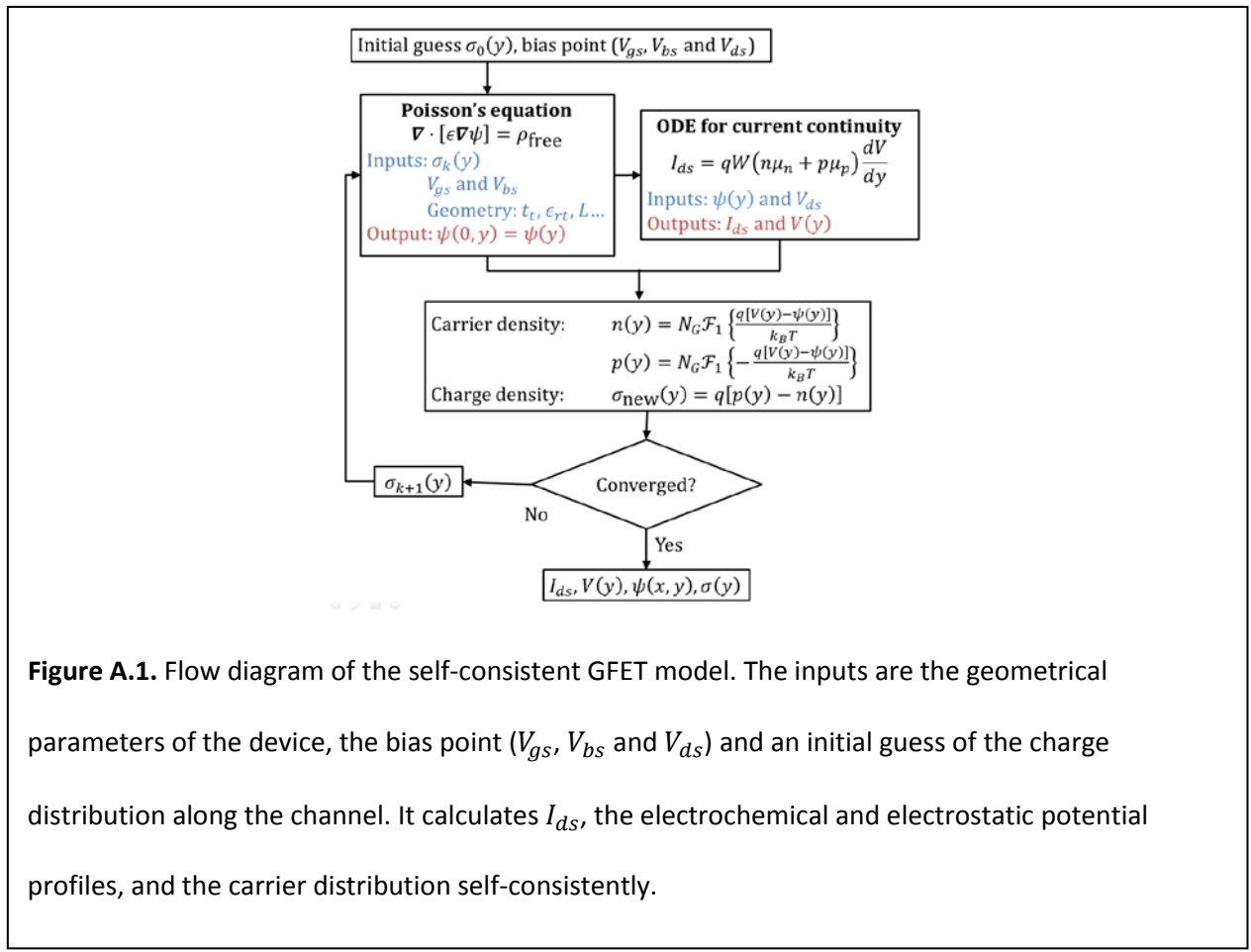

**Figure A.1.** Flow diagram of the self-consistent GFET model. The inputs are the geometrical parameters of the device, the bias point ($V_{gs}$, $V_{bs}$ and $V_{ds}$) and an initial guess of the charge distribution along the channel. It calculates $I_{ds}$, the electrochemical and electrostatic potential profiles, and the carrier distribution self-consistently.

Once $\psi_k(y)$ and $V_k(y)$ have been obtained, the carrier concentrations can be calculated along the channel from equations (A.1) and (A.2) and thus, the charge in the graphene $\sigma_{\text{new},k}(y)$. The algorithm calculates if the solution has converged by comparing $\sigma_{\text{new},k}(y)$ to the guess $\sigma_k(y)$. We define the relative error of the iteration $k$, $\varepsilon_k$, in the following form:

$$\varepsilon_k = \left\{ \frac{\int_0^L [\sigma_{\text{new},k}(y) - \sigma_k(y)]^2 dy}{\int_0^L [\sigma_k(y)]^2 dy} \right\}^{\frac{1}{2}} \tag{A.6}$$

If the algorithm has not converged yet, a new guess must be calculated to reintroduce it in Poisson's equation. This $\sigma_{k+1}(y)$ is obtained from an optimized linear combination of $\sigma_k(y)$ and $\sigma_{\text{new},k}(y)$ of previous iterations, following Pulay's method [56,57]. The procedure to calculate it is thoroughly explained in the supplementary material. This way, the algorithm is able to find the drain current $I_{ds}$, the electrostatic and electrochemical potentials $\psi(x,y)$ and $V(y)$, and carrier distribution, $n(y)$ and $p(y)$, for any given bias point.

# Supplementary material: Short channel effects in graphene-based field effect transistors targeting radio-frequency applications

Pedro C Feijoo, David Jiménez, Xavier Cartoixà

Departament d'Enginyeria Electrònica, Escola d'Enginyeria, Universitat Autònoma de Barcelona, Campus UAB, E-08193 Bellaterra, Spain

**E-Mail:** PedroCarlos.Feijoo@uab.cat

## 1. Pulay mixing

As stated in the main text, the initial guess of the iteration $k + 1$, $\sigma_{k+1}(y)$, is obtained from an optimized linear combination of $\sigma_k(y)$ and $\sigma_{\text{new},k}(y)$ from the $N$ previous iterations by means of Pulay mixing [1,2]. This method was developed to accelerate the convergence of large systems of equations. First, the residual $\Delta\sigma_k$ must be defined.

$$\Delta\sigma_k(y) = \sigma_{\text{new},k}(y) - \sigma_k(y) \tag{S1}$$

We must recall that the initial guess $\sigma_k(y)$ is introduced in equations (1) and (2) in order to obtain the potential profiles $\psi_k(y)$ and $V_k(y)$. From these profiles, equations (A.1) and (A.2) allow us to obtain $\sigma_{\text{new},k}(y)$ (see figure A.1). For the $k^{\text{th}}$ iteration we must build a symmetrical $N \times N$ matrix according to this equation:

$$A_{ij} = \int_0^L \Delta\sigma_i(y)\Delta\sigma_j(y)dy \tag{S2}$$

The indexes $i$ and $j$ run the $N$ iterations that precede iteration $k + 1$, that is, $i,j = k - (N - 1), \ldots, k$. The linear combination of the previous iterations is obtained from the following equation:

$$\sigma_{k+1}(y) = \sum_{l=k-(N-1)}^{k} c_l[\sigma_l(y) + \alpha\Delta\sigma_l(y)] \tag{S3}$$

where $\alpha$ is a parameter of the method, fulfilling $0 < \alpha < 1$. The coefficients $c_l$ are obtained from the inverse of the matrix $A$ according to:

$$c_l = \frac{\sum_j A_{lj}^{-1}}{\sum_i \sum_j A_{ij}^{-1}} \tag{S4}$$

The new initial guess is introduced again in equations (1) and (2). The loop continues until the residual $\varepsilon_k$, defined in equation A.6, decreases below the tolerance $\varepsilon_{\min}$. Parameters of the method used in this work are summarized in Table S1.

**Table S1.** Pulay mixing parameters.

| Parameter | Value |
|---|---|
| $\alpha$ | 0.1 |
| $N$ | 13 |
| $\varepsilon_{\min}$ | $10^{-3}$ |

## 2. Small-signal parameters of the GFET

Figure S1 (a) depicts the typical values of the transconductance $g_m$ for the reference device (Table 1) for a 100 nm channel length. This magnitude is represented as a function of $V_{gs}$ for two different $V_{ds}$ (0.1 and 0.6 V). In the same way, output conductance $g_d$ is represented in figure S1 (b). Figure S1 (c) shows both the source and drain capacitances ($C_{gs}$ and $C_{gd}$, respectively). They are compared with the top gate oxide capacitance $C_t$, as calculated by $C_t = WL\epsilon_{rt}\epsilon_0/t_t$.

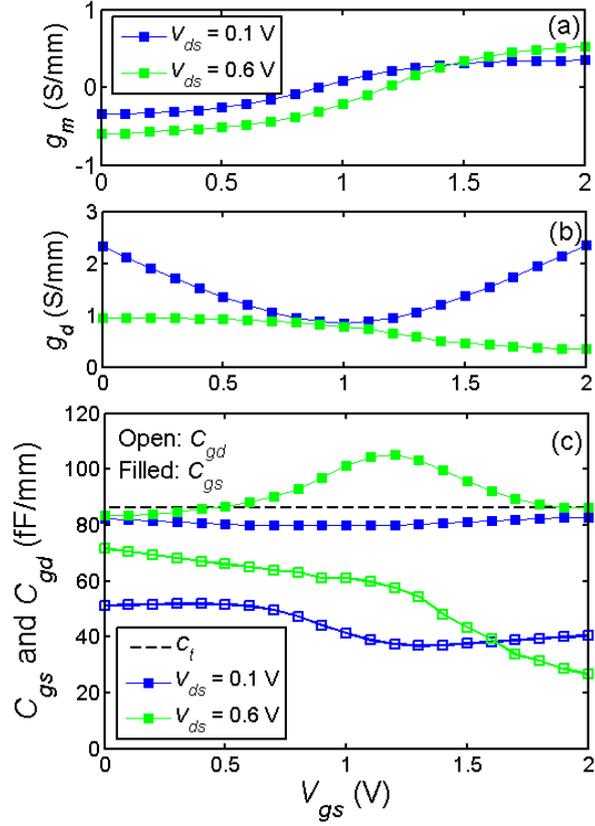

**Figure S1.** (a) Transconductance ($g_m$) as a function of $V_{gs}$ for $V_{ds}$ = 0.1 V and 0.6 V. (b) Output conductance ($g_d$). (c) Capacitances $C_{gd}$ (open symbols) and $C_{gs}$ (filled symbols) as a function of $V_{gs}$. The dashed line correspond to the top oxide capacitance $C_t$. The reference GFET with parameters given in Table 1 was considered with $L$ = 100 nm.

## 3. Election of the bias point to determine $f_T$, $f_{\max}$ and $A_v$

Figure S2 shows the cutoff frequencies as a function of the bias point for the reference device with a channel length of 100 nm. Filled symbols correspond to the intrinsic cutoff frequency $f_{T,i}$, while open symbols to the extrinsic cutoff frequencies $f_{T,x}$, that is, taking account of the source and drain series resistances. Cutoff frequencies go through a minimum when $g_m = 0$, which happens at Dirac's voltage. For each value of $V_{ds}$, the $f_{T,i}$ and $f_{T,x}$ represented in figure 8 (a) and (b), respectively, are the maximum frequencies given in figure S2. The frequency $f_{\max}$ of figure 8 (c) are obtained by the same procedure.

Regarding the voltage gain $A_v$, the extracted values, represented in figure 8 (d), correspond to the maximum value at $V_{gs}$ = 2 V.

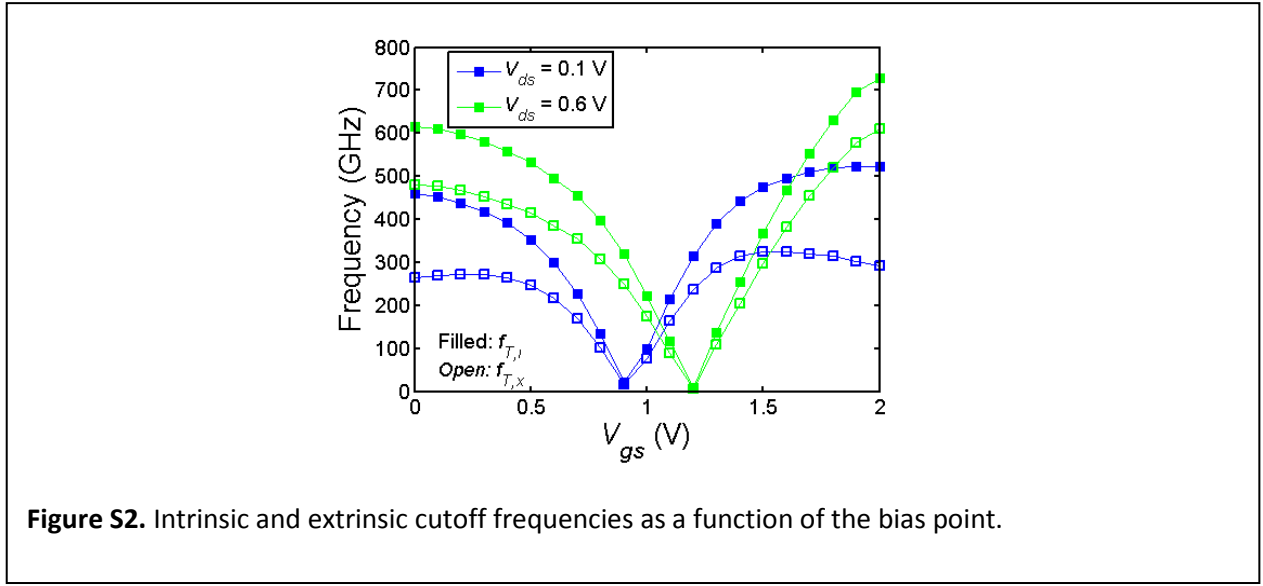

**Figure S2.** Intrinsic and extrinsic cutoff frequencies as a function of the bias point.

## 4. Capacitance, transconductance and output conductance scaling

In order to get a deeper insight into the scaling behavior of GFET, we have represented in figure S3 the gate capacitance $C_g = C_{gs} + C_{gd}$ (a), $g_m$ (b), and $g_d$ (c), as a function of the channel length. All of them are represented for drain voltages of 0.1 V (green symbols) and 0.6 V (blue symbols) at the same gate biases where $f_{T,i}$ is extracted. The gate capacitance values scales linearly in the whole range, and it presents a value that is very close to the top gate oxide capacitance $C_t$. For long-channel (above 1 µm) transistors, $g_m$ approximately scales as $1/L$. By contrast, the increase in $g_d$ results in a poorer saturation of the GFET. The $1/L$ trend continues for $g_m$ at short lengths when the bias is as low as 25 mV, as it can be seen in red symbols of figure S3 (b), but it deviates for higher biases. Downscaling also produces a degradation of $g_d$. SCE are especially important at high bias (0.6 V): the transconductance saturates and even decreases for $L <$ 100 nm. At low bias, the output conductance saturates as $L$ scales, while it further increases at high bias.

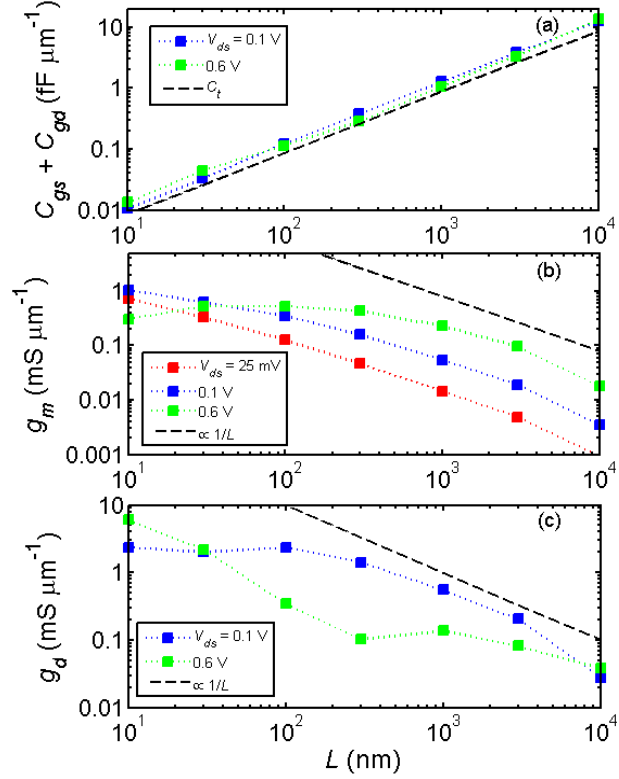

**Figure S3.** Behavior of the (a) gate capacitance, (b) transconductance, and (c) output conductance as channel length scales for drain voltages of 0.1 and 0.6 V. Transconductance is also represented for 25 mV. Gate capacitance ($C_g = C_{gs} + C_{gd}$) scales as $C_t$. In (b) and (c) the dashed lines represent an arbitrary function that scales as $1/L$ for the sake of comparison.

## 5. EOT influence on long- and short-channel devices

In the main text, we state that the current characteristics of a short-channel GFET are influenced not only by the gate capacitance but also by the top oxide thickness and permittivity value themselves, because of SCE. Transfer characteristics are shown in figure S4, where we have compared two GFETs with the same EOT but with different oxide thickness and permittivity for both long channel (a) and short channel (b) cases. The long-channel transistor maintains its characteristics since the gate capacitance is kept constant. However, the drain current is influenced in a dissimilar way by either $t_t$ or $\epsilon_{rt}$ in the short-channel device.

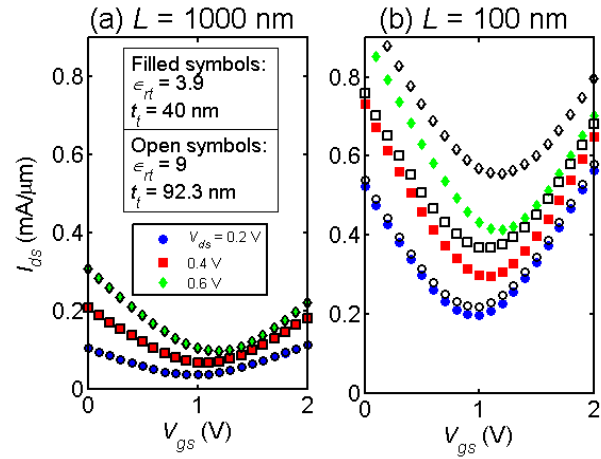

**Figure S4.** GFET transfer characteristics assuming channel lengths of 1 μm (a) and 100 nm (b), respectively. Both filled and open symbols represent a device with the same EOT, although the top oxide thickness and permittivity differ.